\documentclass[12pt]{article}
\usepackage{a4wide,epsfig,psfrag,amsmath}

\newcommand{\agt}{\,\rlap{\lower 3.5 pt \hbox{$\mathchar \sim$}} \raise 1pt
 \hbox {$>$}\,}
\newcommand{\alt}{\,\rlap{\lower 3.5 pt \hbox{$\mathchar \sim$}} \raise 1pt
 \hbox {$<$}\,}

\catcode`@=11
\newcount\@tempcntc
\def\@citex[#1]#2{\if@filesw\immediate\write\@auxout{\string\citation{#2}}\fi
  \@tempcnta\z@\@tempcntb\m@ne\def\@citea{}\@cite{\@for\@citeb:=#2\do
    {\@ifundefined
       {b@\@citeb}{\@citeo\@tempcntb\m@ne\@citea\def\@citea{,}{\bf ?}\@warning
       {Citation `\@citeb' on page \thepage \space undefined}}%
    {\setbox\z@\hbox{\global\@tempcntc0\csname b@\@citeb\endcsname\relax}%
     \ifnum\@tempcntc=\z@ \@citeo\@tempcntb\m@ne
       \@citea\def\@citea{,}\hbox{\csname b@\@citeb\endcsname}%
     \else
      \advance\@tempcntb\@ne
      \ifnum\@tempcntb=\@tempcntc
      \else\advance\@tempcntb\m@ne\@citeo
      \@tempcnta\@tempcntc\@tempcntb\@tempcntc\fi\fi}}\@citeo}{#1}}
\def\@citeo{\ifnum\@tempcnta>\@tempcntb\else\@citea\def\@citea{,}%
  \ifnum\@tempcnta=\@tempcntb\the\@tempcnta\else
   {\advance\@tempcnta\@ne\ifnum\@tempcnta=\@tempcntb \else \def\@citea{--}\fi
    \advance\@tempcnta\m@ne\the\@tempcnta\@citea\the\@tempcntb}\fi\fi}
\catcode`@=12

\begin{document}
\allowdisplaybreaks

\title{
\vskip-3cm{\baselineskip14pt
\centerline{\normalsize DESY 04-224\hfill ISSN 0418-9833}
\centerline{\normalsize hep-ph/0411300\hfill}
\centerline{\normalsize November 2004\hfill}}
\vskip1.5cm
Inclusive Production of Single Hadrons with Finite Transverse Momenta
in Deep-Inelastic Scattering at Next-to-Leading Order}

\author{B.A. Kniehl, G. Kramer, M. Maniatis\\
{\normalsize II. Institut f\"ur Theoretische Physik, Universit\"at Hamburg,}\\
{\normalsize Luruper Chaussee 149, 22761 Hamburg, Germany}}

\date{}

\maketitle

\begin{abstract}
We calculate the cross section for the inclusive production of single hadrons
with finite transverse momenta in deep-inelastic scattering at next-to-leading
order (NLO), i.e.\ through ${\cal O}(\alpha_s^2)$, in the parton model of QCD 
endowed with non-perturbative parton distribution functions (PDFs) and
fragmentation functions (FFs).
The NLO correction is found to produce a sizeable enhancement in cross
section, of up to one order of magnitude, bringing the theoretical prediction
to good agreement with recent measurements for neutral pions and charged
hadrons at DESY HERA.
This provides a useful test for the universality and the scaling violations of
the FFs predicted by the factorization theorem.
Such comparisons can also be used to constrain the gluon PDF of the proton.

\medskip

\end{abstract}

\newpage

\section{Introduction}
\label{sec:int}
 
The predictive power of the parton model of quantum chromodynamics (QCD) lies
in the factorization theorem.
In deep-inelastic scattering (DIS), factorization in short- and long-distance
parts allows us to describe the observed cross sections of inclusive hadron
production as a convolution of the partonic cross sections with
non-perturbative parton density functions (PDFs) and fragmentation functions
(FFs) \cite{Collins:gx}.
Single-hadron inclusive production in electron-proton DIS,
\begin{equation}
\label{eqeP}
e^-(k) + p(P) \to e^-(k^\prime) + h(p) + X,
\end{equation}
occurs partonically already in the absence of strong interactions, at 
${\mathcal O}(\alpha_s^0)$, where $\alpha_s$ is the strong-coupling constant,
when one parton of the proton (a quark) interacts with the lepton current and
fragments into the hadron $h$ (na\"\i ve parton model).
If the virtuality $Q^2=-q^2$ of the four-momentum transfer $q=k^\prime-k$
satisfies $Q^2\ll m_Z^2$, where $m_Z$ denotes the $Z$-boson mass, then
process~(\ref{eqeP}) is essentially mediated by a virtual photon ($\gamma^*$),
while the contribution from $Z$-boson exchange is negligible.
In the following, this is the situation we are interested in.

Since we are interested in perturbative QCD effects, we require the hadron to
carry non-vanishing transverse momentum ($p_T^*$) in the centre-of-mass (c.m.)
frame of the virtual photon and the incoming proton.
At leading order (LO), the corresponding partonic subprocesses thus contain
two partons in the final state, one of which fragments into the hadron, while
the other one balances the transverse momentum.
At next-to-leading order (NLO), three-parton final states contribute to the
real correction, while the virtual correction arises from one-loop diagrams
with two final-state partons.

The investigation of single-hadron production is interesting for several
reasons.
First of all, it provides a test of perturbative QCD and of factorization.
Apart from the partonic cross sections obtainable from perturbative QCD,  
the theoretical predictions essentially depend on universal PDFs and FFs.
In particular, the FFs, which are fitted to electron-positron-annihilation
data, may be tested with regard to their universality.
Furthermore, the theoretical predictions allow for a direct comparison with
experimental data, without resorting to any kind of Monte Carlo model to
simulate the hadronization of the outgoing partons. 
Thus, we may expect very meaningful results.
Moreover, the theoretical predictions are directly sensitive to the gluon PDF
of the proton with the potential to constrain the latter.

On the experimental side, precise data were collected by the H1
\cite{Adloff:1996dy,Adloff:1999zx,Aktas:2004rb} and ZEUS
\cite{Derrick:1995xg,Breitweg:1999nt} Collaborations at the $ep$ collider HERA
at DESY.
They refer to $\pi^0$ mesons in the forward region
\cite{Adloff:1999zx,Aktas:2004rb}, with small angles with respect to the
proton remnant, and to charged hadrons
\cite{Adloff:1996dy,Derrick:1995xg,Breitweg:1999nt}.

More than 25 years ago, the cross section of process~(\ref{eqeP}) with
finite transverse momentum of the hadron $h$ was calculated by M\'endez
\cite{Mendez:zx} at LO, to ${\mathcal O}(\alpha_s)$.
Since QCD corrections are typically large and we are confronted with precise 
experimental data, it is desirable to compare these data with predictions of
at least NLO accuracy, including the terms of ${\mathcal O}(\alpha_s^2)$.
For this purpose, a first NLO QCD prediction was computed in
Ref.~\cite{Buettner}, neglecting the longitudinal degrees of freedom of the
virtual photon.

The theoretical description can be rendered more reliable by resumming the
leading logarithmic contributions of the perturbation expansion. 
In this sense, the Balitsky-Fadin-Kuraev-Lipatov (BFKL) \cite{BFKL} and
Dokshitser-Gribov-Lipatov-Altarelli-Parisi (DGLAP) \cite{DGLAP} equations
resum at LO the $(\alpha_s\ln(1/x_B))^n$ and $(\alpha_s\ln(Q^2/Q_0^2))^n$
contributions, respectively, where $x_B$ is the Bjorken variable and $Q_0$ is
the cut-off scale for the perturbative evolution.
Disregarding the fact that these resummations are just approximations, they
evidently fail in the kinematic regions of large $x_B$ values and small $Q^2$
values, respectively.

In this paper, we perform a full NLO QCD calculation, also taking into
account the longitudinal degrees of freedom of the virtual photon.
We encounter ultraviolet (UV) and infrared (IR) singularities, which we all
regularize using dimensional regularization.
In order to overcome the difficulties in connection with the IR singularities
emerging in different parts of the NLO correction, we employ the dipole
subtraction formalism \cite{Catani:1996vz}.
In contrast to the more conventional phase-space slicing method, there is no
need to introduce any unphysical parameter to cut the phase space into soft,
collinear, and hard regions.
Moreover, all cancellations of IR singularities occur before any numerical
phase-space integration is performed. 
We thus conveniently obtain numerically stable predictions.

An independent calculation was recently presented in
Ref.~\cite{Aurenche:2003by}, where the matrix elements of the hard-scattering
processes were adopted from the DISENT program package~\cite{Catani:1996gg}.
In Ref.~\cite{Aurenche:2003by}, the phase-space slicing method was applied to
handle the IR singularities.
Another related work, focusing on fracture functions of the proton, was
published in two parts, related to incoming gluons \cite{Daleo:2003xg} and
quarks \cite{Daleo:2003cf}.

This paper is organized as follows.
In Section~\ref{sec:ana}, we describe our analytical analysis.
The LO result and a specific part of the real NLO correction are relegated to
Appendices~\ref{AppLO} and \ref{AppFurry}, respectively.
In Section~\ref{sec:num}, we present our numerical results.
Our conclusions are summarized in Sec.~\ref{sec:con}.

\section{Analytical analysis}
\label{sec:ana}

\begin{figure}[ht]
\begin{center}
\includegraphics[width=10cm]{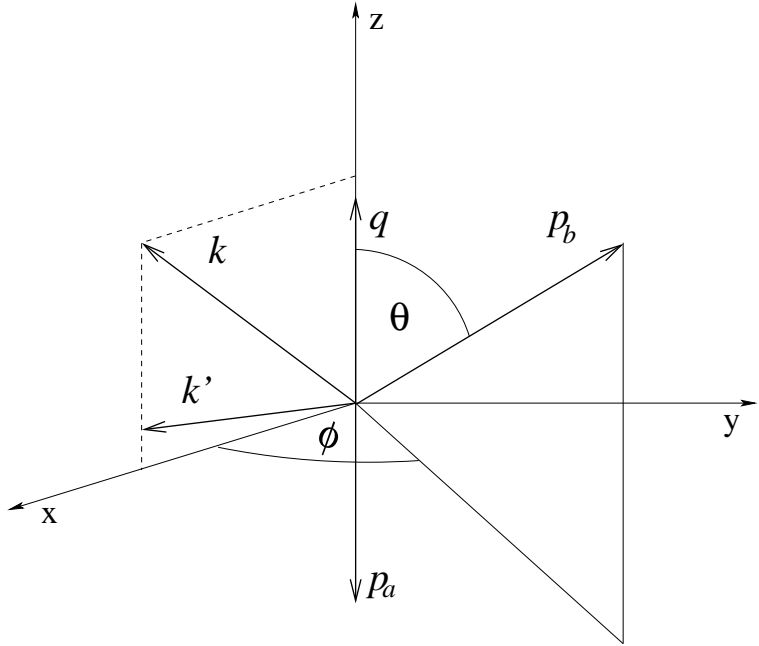}
\caption{C.m.\ frame of the virtual photon and the initial-state parton $a$,
where the three-momenta of the leptons are rotated into the $x$-$z$ plane.}
\label{coordinates}
\end{center}
\end{figure}

According to the factorization theorem \cite{Collins:gx}, the differential
cross section for process~(\ref{eqeP}) is given as a convolution of the
hard-scattering cross sections $d\sigma^{ab}$ with the PDFs $F_a^p$ of the
proton and FFs $D_b^h$ of hadron $h$, as
\begin{equation}
\label{eqconv}
\frac{d^4\sigma^{h}}{d\overline{x}\,dy\,d\overline{z}\,d\phi} =
\sum\limits_{ab}
\int\limits_{\overline{x}}^1 \frac{dx}{x}
\int\limits_{\overline{z}}^1 \frac{dz}{z}
F_a^p(\overline{x}/x,\mu_i) 
\frac{d^4\sigma^{ab}}{dx\,dy\,dz\,d\phi}
D_b^h(\overline{z}/z,\mu_f),
\end{equation}
where $\mu_i$ and $\mu_f$ are the factorization scales related to the initial
and final states and the sum runs over all {\it tagged} initial- and
final-state partons, $a$ and $b$, respectively.
As usual, the dimensionless variables $x$, $y$, and $z$ are defined as
$x=Q^2/(2p_a\!\cdot\!q)$, $y=p_a\!\cdot\!q/p_a\!\cdot\!k$, and
$z=p_a\!\cdot\!p_b/p_a\!\cdot\!q$ with respect to the partonic four-momenta
$p_a$ and $p_b$, and their barred counterparts
$\overline{x}=x_{\rm B}=Q^2/(2P\cdot q)$, $\overline{y}=y=P\cdot q/P\cdot k$,
and $\overline{z}=P\cdot p/P\cdot q$ with respect to the hadronic four-momenta.
We have $Q^2=x_ByS$, where $S=(P+k)^2$ is the square of the $ep$ c.m.\ energy.
It is convenient to describe the kinematics in the c.m.\ frame of the virtual
photon and the incoming parton $a$ as is done in Fig.~\ref{coordinates}, where
we take the three-momentum of the virtual photon to point along the $z$ axis
and the three-momenta of the incoming and scattered electrons to lie in the
$x$-$z$ plane.
Then, the azimuthal angle $\phi$ of the hadron $h$ is enclosed between the
plane spanned by the three-momenta of the incoming and scattered electrons and
the one spanned by those of the virtual photon and the outgoing parton $b$.

The hard-scattering cross sections may be written as contractions of a lepton
tensor $l^{\mu\nu}$ with hadron tensors $H_{\mu\nu}^{ab}$, as
\begin{equation}
\frac{d^4\sigma^{ab}}{dx\,dy\,dz\,d\phi} =
\frac{\alpha^2}{16\pi^2}\,\frac{y}{Q^4} l^{\mu\nu} H_{\mu\nu}^{ab},
\end{equation}
where $\alpha$ is Sommerfeld's fine-structure constant.
If the virtual photon and the initial-state parton are both unpolarized, then
there cannot be any dependence on the azimuthal angle $\phi$.
Integrating over the latter, we find the decomposition
\begin{equation}
\frac{d^3\sigma^{ab}}{dx\,dy\,dz} =
\frac{\alpha^2}{8\pi}
\left(\frac{y^2-2y+2}{2yQ^2}H_T^{ab}+2\frac{y^2-6y+6}{y^3s^2}H_L^{ab}\right),
\label{eqlepton}
\end{equation}
where $H_T^{ab}=-g^{\mu\nu}H_{\mu\nu}^{ab}$,
$H_L^{ab}=p_a^\mu p_a^\nu H_{\mu\nu}^{ab}$, and $s=(p_a+q)^2$.

The partonic subprocesses contributing at LO are
\begin{eqnarray}
\gamma^*+q&\to&q+g,\label{born1}\\
\gamma^*+q&\to&g+q,\label{born2}\\
\gamma^*+g&\to&q+\overline{q},\label{born3}
\end{eqnarray}
where it is understood that the first of the final-state partons is the one
that fragments into the hadron $h$.
Here, $q=q_1,\overline{q}_1,\ldots,q_{n_f},\overline{q}_{n_f}$, where $n_f$ is
the number of active quark flavours, which are ordered according to their
masses, i.e., $q_1=u$, $q_2=d$, $q_3=s$, $q_4=c$, and $q_5=b$, and we identify
$\overline{\overline{q}}=q$.
There are two Feynman diagrams for each of the
processes~(\ref{born1})--(\ref{born3}).
The matrix elements of processes~(\ref{born1})--(\ref{born3}) are interrelated
through crossing symmetry.
Owing to charge-conjugation ($C$) invariance, the counterparts of
processes~(\ref{born1})--(\ref{born3}) with quarks and antiquarks
interchanged yield equal cross sections and do not have to be calculated
separately. 
However, this is not generally true for the PDFs and FFs.
Therefore, we have to explicitly sum over all possible pairings of partons $a$
and $b$ in Eq.~(\ref{eqconv}).
For the reader's convenience, the LO expressions for the Lorentz invariants
$H_T^{ab}$ and $H_L^{ab}$ in Eq.~(\ref{eqlepton}) are listed in
Appendix~\ref{AppLO}.
They are of ${\mathcal O}(\alpha_s)$ and proportional to $e_q^2$, where $e_q$
denotes the electric charge of quark $q$ in units of the positron charge.

In order to determine the NLO correction to the cross section of
processes~(\ref{eqeP}), we have to compute the virtual and real corrections of
${\mathcal O}(\alpha_s^2)$ to the hadron tensors.
We then encounter a rather involved pattern of singularities.
All these singularities are regularized using dimensional regularization with
$D=4-2\epsilon$ space-time dimensions yielding poles in $\epsilon$ in the 
physical limit $D\to4$.
The integrations over the loop four-momenta in the virtual correction lead to
UV and IR singularities, where the IR ones comprise both soft and collinear 
singularities.
All UV singularities are removed through the renormalizations of the
wave functions and the strong-coupling constant.
The remaining soft and collinear singularities cancel partly against
counterparts originating from the phase-space integration of the real
correction.
The remaining collinear poles have to be factorized into the bare PDFs and FFs
so as to render them finite.

The virtual correction is obtained as the interference of the Born and
one-loop matrix elements.
The latter receive contributions from self-energy, triangle, and box diagrams.
These involve two-, three-, and four-point tensor integrals, which are reduced
to scalar integrals via tensor reduction \cite{Passarino:1978jh}.
The scalar integrals contain both UV and IR singularities.
They are computed analytically in dimensional regularization. 
Our analytic expressions for the contractions of the resulting hadron tensors
with $g^{\mu\nu}$ agree with the literature \cite{Graudenz:1993tg}. 
The virtual correction is renormalized in the modified minimal-subtraction
($\overline{\text{MS}}$) scheme and thus UV finite.

The partonic subprocesses contributing to the real correction read
\begin{eqnarray}
\gamma^*+q&\to&q+g+g,\label{real1}\\
\gamma^*+q&\to&g+q+g,\label{real2}\\
\gamma^*+g&\to&q+\overline{q}+g,\label{real3}\\
\gamma^*+g&\to&g+q+\overline{q},\label{real4}\\
\gamma^*+q&\to&q+q+\overline{q},\label{real5}\\
\gamma^*+q&\to&\overline{q}+q+q,\label{real6}\\
\gamma^*+q&\to&q+q^\prime+\overline{q}^\prime,\label{real7}\\
\gamma^*+q&\to&q^\prime+\overline{q}^\prime+q,\label{real8}
\end{eqnarray}
where $q,q^\prime=q_1,\overline{q}_1,\ldots,q_{n_f},\overline{q}_{n_f}$ with
$q\ne q^\prime$.
As in processes~(\ref{born1})--(\ref{born3}), the first partons in the final
states of processes~(\ref{real1})--(\ref{real8}) are taken to fragment into
the hadron $h$.
The order in which the residual final-state partons appear is irrelevant.
There are eight Feynman diagrams for each of the
processes~(\ref{real1})--(\ref{real6}) and four ones for each of the
processes~(\ref{real7}) and (\ref{real8}).
Crossing symmetry interrelates the matrix elements of
processes~(\ref{real1})--(\ref{real4}), those of
processes~(\ref{real5})--(\ref{real6}), and those of 
processes~(\ref{real7})--(\ref{real8}).
The cross sections of processes~(\ref{real1})--(\ref{real8}) are of
${\mathcal O}(\alpha_s^2)$.
Those of processes~(\ref{real1})--(\ref{real6}) are proportional to $e_q^2$,
while, at first sight, those of processes~(\ref{real7})--(\ref{real8}) contain
pieces proportional to $e_q^2$, $e_qe_{q^\prime}$, and $e_{q^\prime}^2$.
However, in the case of process~(\ref{real7}), the piece proportional to
$e_qe_{q^\prime}$ vanishes by Furry's theorem, as is explained below.
The squared matrix elements of processes~(\ref{real1})--(\ref{real4}) involve
one quark trace, those of processes~(\ref{real5}) and (\ref{real6}) contain
pieces with one or two quark traces, and those of processes~(\ref{real7}) and 
(\ref{real8}) involve two quark traces.
Due to $C$ invariance, the counterparts of
processes~(\ref{real1})--(\ref{real8}) with quarks and antiquarks interchanged
yield equal cross sections.
Notice that, in the case of process~(\ref{real8}), we have to distinguish
between the case where the tagged quarks $q$ and $q^\prime$ are both particles
or anti-particles and the case where one is a particle and the other one is an
anti-particle. 
Processes~(\ref{real1}) and (\ref{real6}) each contain two identical untagged
partons in the final state, so that their cross sections receive a statistical
factor of 1/2 to avoid double counting in the phase space integration.
We derived the matrix elements of processes~(\ref{real1})--(\ref{real8}) in
two steps.
First, we calculated the ones of the corresponding processes of $e^+e^-$
annihilation via a virtual photon, which may also be found in
Ref.~\cite{Ellis:1980wv}.
Then, we employed crossing symmetry.
The squared matrix elements of processes~(\ref{real1})--(\ref{real8}),
excluding the Furry terms discussed below, are also implemented in the DISENT
program package \cite{Catani:1996gg}.
Performing a numerical comparison with the latter, we find agreement.

\begin{figure}[ht]
\begin{center}
\includegraphics[width=10cm]{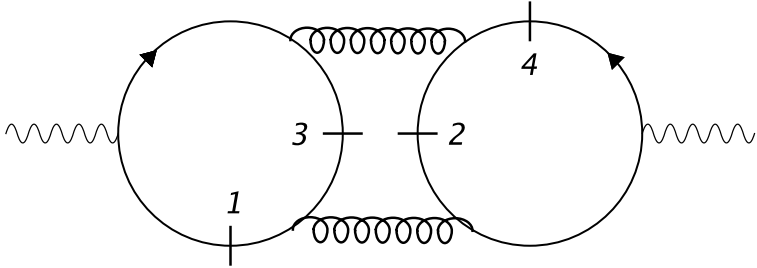}
\caption{Cut-diagram involving two fermion traces each coupled to three gauge
bosons, as illustrated in Ref.~\cite{Ellis:1980wv}.
The cut proceeds along the numbered ticks representing the four on-shell
quarks. 
If the charges of both quark loops are tagged, there is no Furry cancellation
with an analogous diagram with one fermion-number flow flipped.}
\label{Furry}
\end{center}
\end{figure}

In Ref.~\cite{Ellis:1980wv}, the squared matrix elements, which may be
visualized as cut diagrams, are classified with respect to colour factors.
One specific class, called {\it F terms}, contains all cut diagrams with two
fermion loops, which are both coupled to three vector bosons, namely to one
photon and two gluons.
This class constitutes a gauge-parameter-independent subset of the NLO
correction.
The cut diagram of one specific member of this class is shown in
Fig.~\ref{Furry}, where the on-shell quarks are indicated by numbers.
As was noticed in Ref.~\cite{Ellis:1980wv}, by Furry's theorem, each cut
diagram within this class exactly cancels against one counterpart in which one
fermion-number flow is reversed if the on-shell quarks associated with the
loop whose fermion-number flow is reversed are not tagged in the
experiment, i.e.\ if the three-momenta of these quarks are integrated over.
This argument is also true in the case where only one fermion charge is
identified, for instance in single-hadron production by $e^+e^-$ annihilation,
since there is still a counterpart diagram where the other fermion-number flow
is reversed. 
In our case, there are two tagged partons, one coming from the proton and one
fragmenting into the hadron $h$.
Suppose the two tagged partons are the quarks 1 and 2 in the cut diagram of
Fig.~\ref{Furry}.
This situation can occur for processes~(\ref{real5}) and (\ref{real6}), which
involve only one quark flavour, and for process~(\ref{real8}), which involves
two different quark flavours.
Then, the Furry cancellation is impeded because there is no counterpart 
diagram.
Thus, we are not allowed to omit this class of cut diagrams in our
calculation.
The corresponding squared matrix elements are listed in
Appendix~\ref{AppFurry}.

The differential cross sections of processes~(\ref{real1})--(\ref{real8}) have
to be integrated over the three-momenta of the second and third final-state
partons keeping the three-momentum of the first one fixed.
Performing the phase-space integrations, we encounter IR singularities of the
soft and/or collinear types, which, for consistency with the virtual
correction, must be extracted using dimensional regularization.
It is convenient to do this by means of the dipole subtraction formalism
\cite{Catani:1996vz}.
The general idea of this formalism is to subtract from the contribution to the
real correction due to a given partonic subprocess some artificial counterterm
which has the same point-wise IR-singular behaviour in $D$ space-time
dimensions as the considered part of the real correction itself. 
Thus, the limit $\epsilon\to0$ can be performed, and the phase space
integration can be evaluated numerically in four dimensions.
The artificial counterterm is constructed in such a way that it can be
integrated over the one-parton subspace analytically leading to poles in
$\epsilon$.
Adding the terms thus constructed to the virtual correction, the IR
singularities of the latter are cancelled analytically.
In the present case, where the three-momenta of two tagged partons need to be
kept fixed, additional, more complicated artificial counterterms appear than
in situations where only one parton is tagged, such as inclusive jet 
production.
A technical advantage of the dipole subtraction method compared to the phase
space slicing method is that all IR singularities cancel before any numerical
integration is performed.
Furthermore, there is no need to introduce a slicing parameter to separate
soft and/or collinear phase space regions from the remaining hard region, 
which needs to be tuned in order to obtain a numerically stable result.
For the factorization of the collinear singularities associated with the
tagged partons, we choose the $\overline{\text{MS}}$ scheme.
In turn, we have to employ PDFs and FFs which are defined in the same scheme.

Finally, we end up with two contributions, the real correction with the
artificial counterterms subtracted and the virtual correction with the
integrated artificial counterterms included, which are both finite in the
physical limit $\epsilon\to0$ and can be integrated over their three- and
two-particle phase spaces, respectively, in three spacial dimensions.
These integrations are performed numerically using a custom-made
{\tt C}$^{++}$ routine.
On the other hand, all algebraic calculations are executed with help of the
symbolic-manipulation package {\tt FORM} \cite{Vermaseren:2000nd}. 

\section{Numerical results}
\label{sec:num}

We are now in a position to present our numerical results for the cross
section of single-hadron inclusive production in $ep$ DIS.
We start by specifying our input.
We work in the $\overline{\text{MS}}$ renormalization and factorization scheme
with $n_f=5$ massless quark flavours.
At NLO (LO), we employ set CTEQ6M (CTEQ6L1) of proton PDFs by the Coordinated
Theoretical-Experimental Project on QCD (CTEQ) \cite{Pumplin:2002vw}, the NLO
(LO) set of FFs for light charged hadrons ($\pi^\pm$, $K^\pm$, and
$p/\overline{p}$) by Kniehl, Kramer, and P\"otter (KKP) \cite{Kniehl:2000fe},
and the two-loop (one-loop) formula for $\alpha_s^{(n_f)}(\mu_r)$ with
asymptotic scale parameter $\Lambda_{\rm QCD}^{(5)}=226$~MeV (165~MeV)
\cite{Pumplin:2002vw}.
This value is compatible with the result
$\Lambda_{\rm QCD}^{(5)}=(213\pm80)$~MeV ($(88\pm41)$~MeV) determined in
Ref.~\cite{Kniehl:2000cr}.
We approximate the $\pi^0$ FFs as
\begin{equation}
D_a^{\pi^0}(x,\mu_f)=\frac{1}{2}D_a^{\pi^\pm}(x,\mu_f),
\end{equation}
where $D_a^{\pi^\pm}$ refers to the sum of the $\pi^+$ and $\pi^-$ mesons,
which is supported by LEP1 data of hadronic $Z^0$-boson decays
\cite{Adam:1995rf}.
Furthermore, we assume the charged hadrons to be exhausted by the charged 
pions, charged kaons, protons, and antiprotons, viz
\begin{equation}
D_a^{h^\pm}(x,\mu_f)=D_a^{\pi^\pm}(x,\mu_f)+D_a^{K^\pm}(x,\mu_f)
+D_a^{p/\overline{p}}(x,\mu_f).
\end{equation}
For simplicity, we identify the renormalization scale $\mu_r$ and the initial-
and final-state factorization scales, $\mu_i$ and $\mu_f$, respectively, and
relate them to the characteristic dimensionful variables $Q^2$ and $p_T^*$ by
setting $\mu_r^2=\mu_i^2=\mu_f^2=\xi[Q^2+(p_T^*)^2]/2$, where $\xi$ is a
dimensionless parameter of order unity introduced to estimate the theoretical
uncertainty due to unphysical-scale variations.
As usual, we consider variations of $\xi$ between 1/2 and 2 about the default 
value 1.

\begin{figure}[t]
\begin{center}
\includegraphics[width=14cm,angle=270]{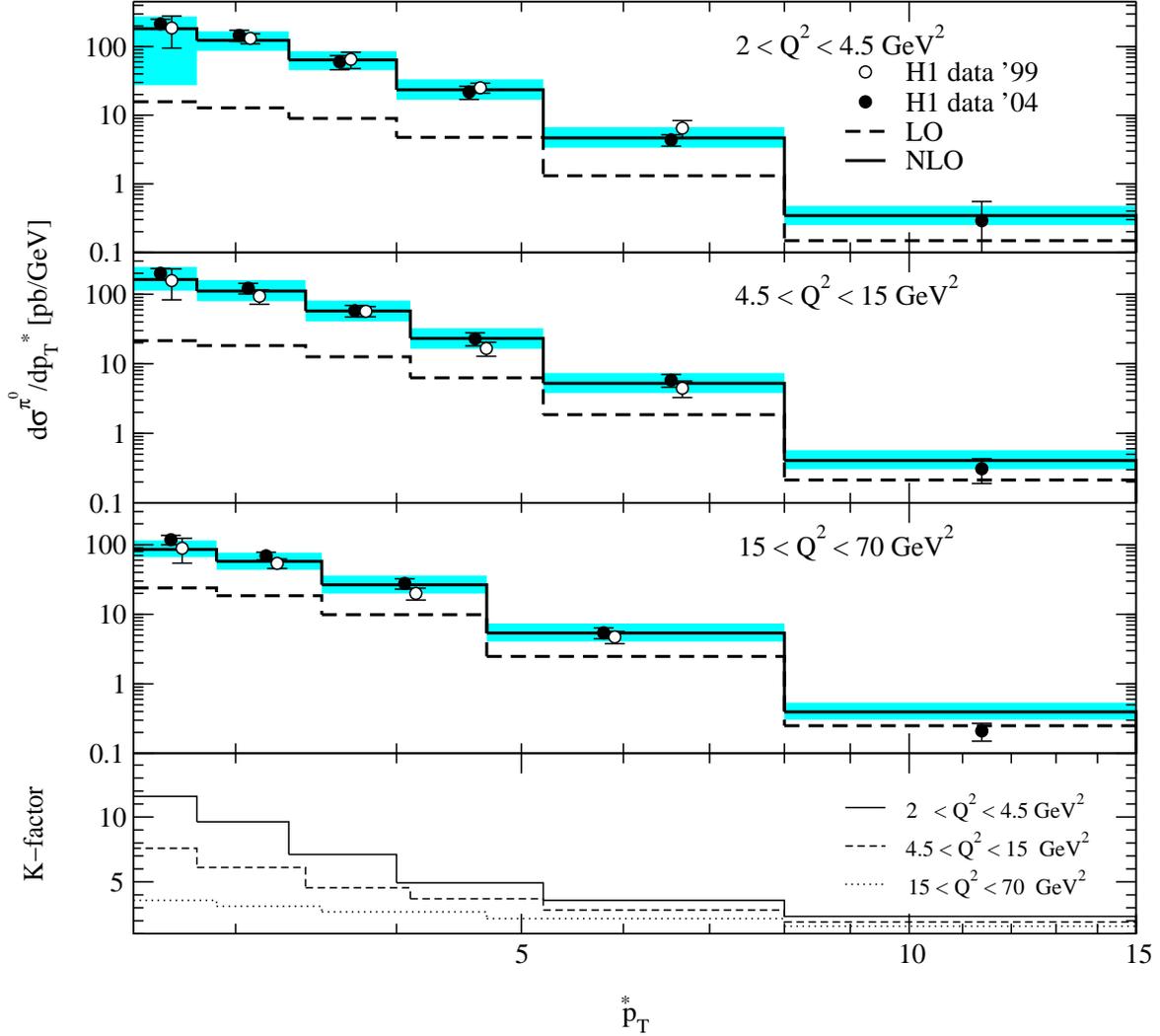}
\caption{Differential cross section $d\sigma^{\pi^0}/dp_T^*$ (in pb/GeV) of
$e^+p\to e^+\pi^0+X$ in DIS with $0.1<y<0.6$ and $2<Q^2<4.5$~GeV$^2$ (first
frame), $4.5<Q^2<15$~GeV$^2$ (second frame), or $15<Q^2<70$~GeV$^2$ (third
frame) at HERA with $E_e=27.6$~GeV and $E_p=820$~GeV for $\pi^0$ mesons with
$5^\circ<\theta<25^\circ$ and $x_E>0.01$.
H1 data from Refs.~\cite{Adloff:1999zx} (open circles) and \cite{Aktas:2004rb}
(solid circles) are compared with our default LO (dashed histograms) and NLO
(solid histograms) predictions including theoretical uncertainties due to
$\xi$ variation (shaded bands).
The $K$ factors (fourth frame) are also shown.}
\label{fig:p}
\end{center}
\end{figure}

\begin{figure}[t]
\begin{center}
\includegraphics[width=14cm,angle=270]{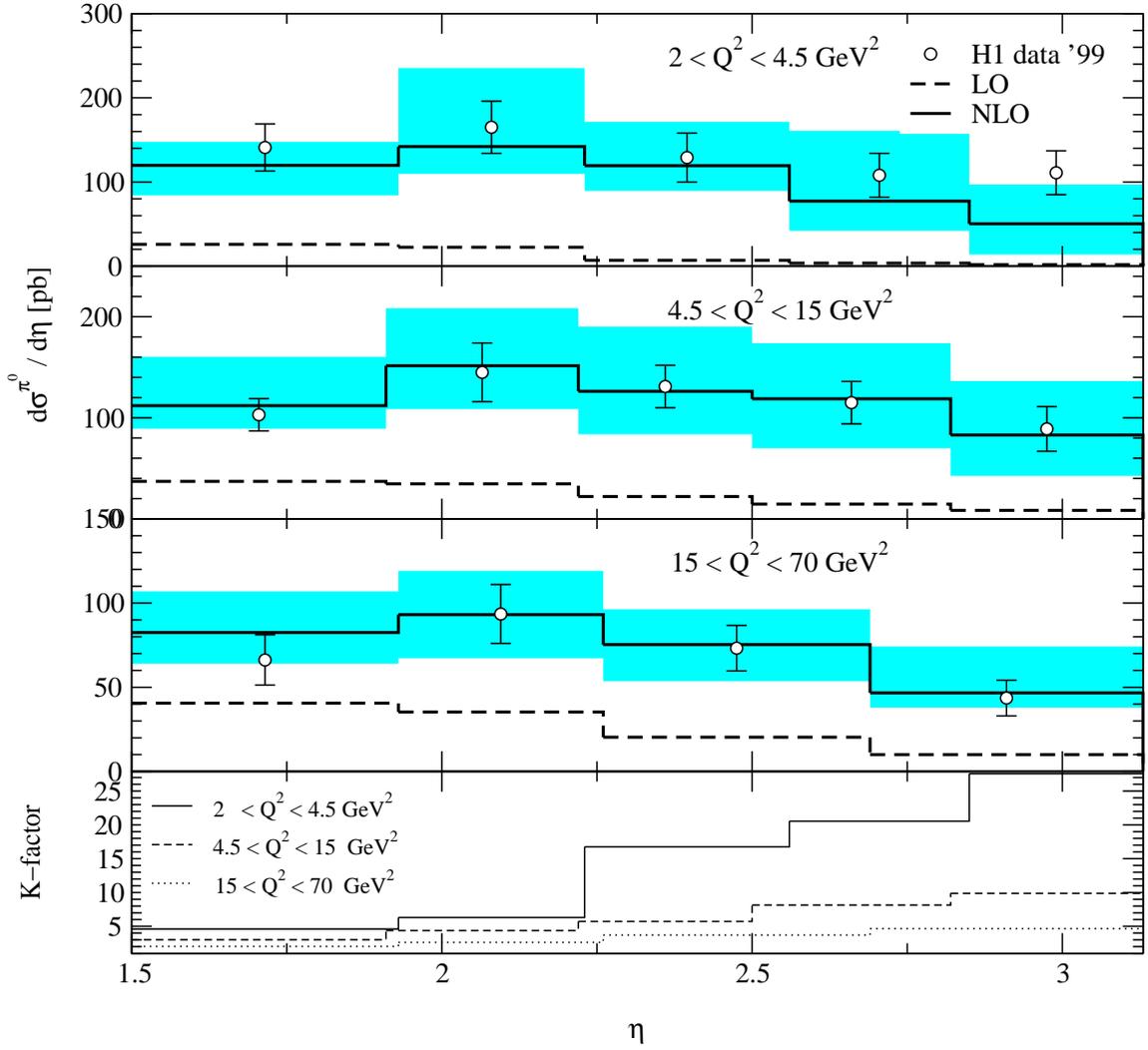}
\caption{Same as in Fig.~\ref{fig:p}, but for $d\sigma^{\pi^0}/d\eta$ (in pb)
with $p_T^*>2.5$~GeV.}
\label{fig:e}
\end{center}
\end{figure}

\begin{figure}[t]
\begin{center}
\includegraphics[width=14cm]{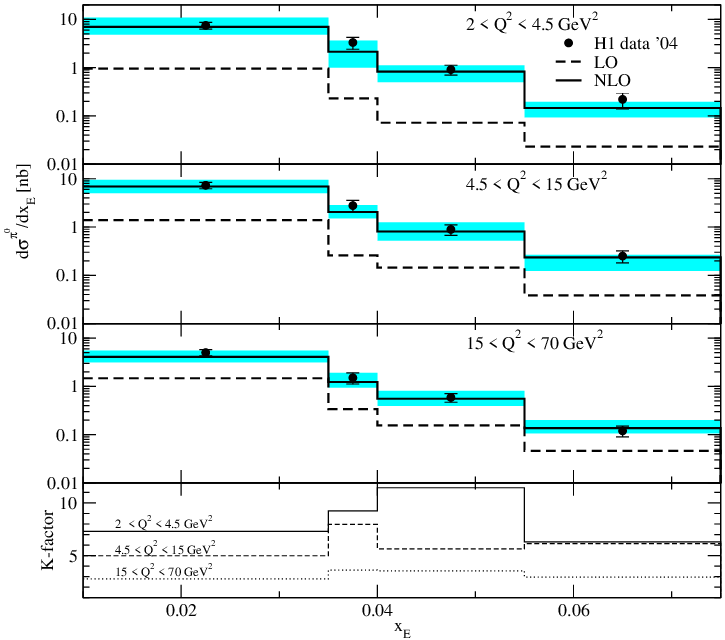}
\caption{(a) Same as in Fig.~\ref{fig:p}, but for $d\sigma^{\pi^0}/dx_E$ (in
nb) with $p_T^*>2.5$~GeV.}
\label{fig:x}
\end{center}
\end{figure}

\begin{figure}[t]
\begin{center}
\includegraphics[width=14cm,angle=270]{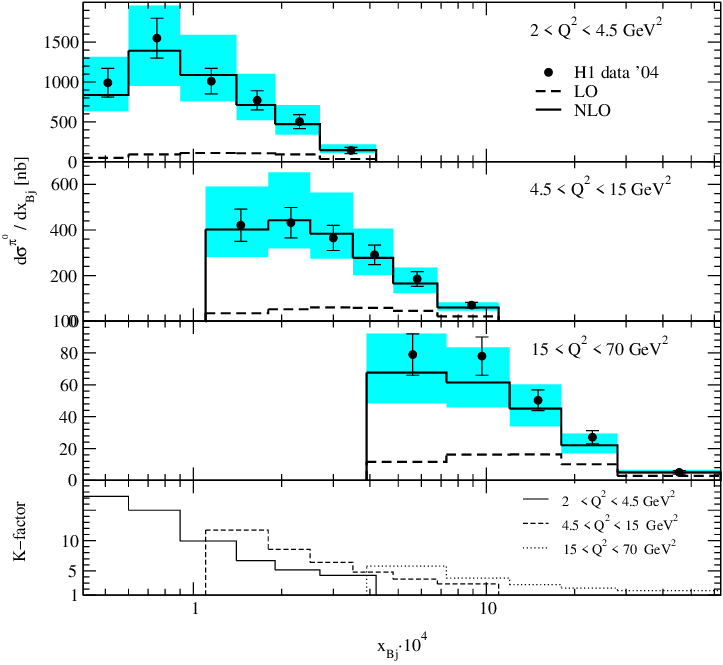}
\caption{(a) Same as in Fig.~\ref{fig:p}, but for $d\sigma^{\pi^0}/dx_B$ (in
nb) with $p_T^*>2.5$~GeV.}
\label{fig:b}
\end{center}
\end{figure}

\begin{figure}[t]
\begin{center}
\includegraphics[width=14cm]{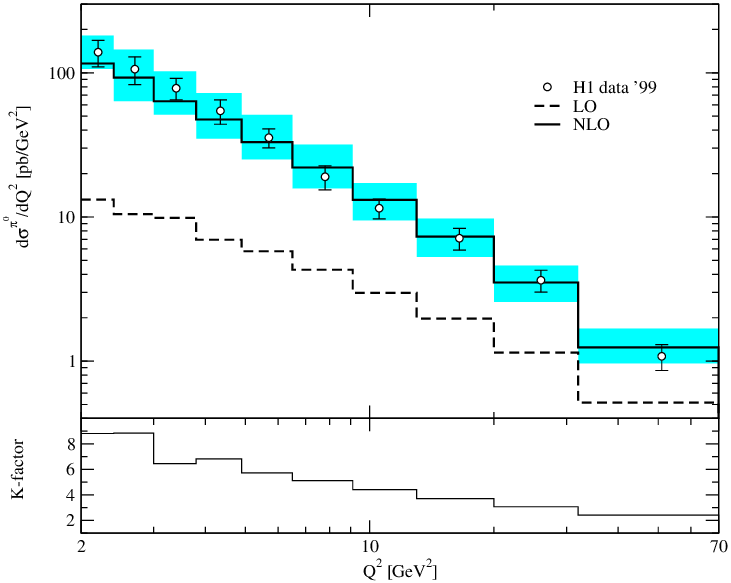}
\caption{Same as in Fig.~\ref{fig:p}, but for $d\sigma^{\pi^0}/dQ^2$ (in
pb/GeV$^2$) with $p_T^*>2.5$~GeV.}
\label{fig:q}
\end{center}
\end{figure}

We now compare our theoretical predictions with HERA data on $\pi^0$ mesons in
the forward region from the H1 Collaboration \cite{Adloff:1999zx,Aktas:2004rb}
and on charged hadrons in the current-jet region from the ZEUS Collaboration
\cite{Derrick:1995xg}.
We start by discussing the H1 data \cite{Adloff:1999zx,Aktas:2004rb}, which
were taken in DIS of positrons with energy $E_e=27.6$~GeV on protons with
energy $E_p=820$~GeV in the laboratory frame, so that
$\sqrt S=2\sqrt{E_eE_p}=301$~GeV, during the running periods 1996 and
1996/1997, and correspond to integrated luminosities of 5.8 and
21.2~pb$^{-1}$, respectively.
In Refs.~\cite{Adloff:1999zx,Aktas:2004rb}, the $\pi^0$ mesons were described
by their transverse momentum $p_T^*$ in the $\gamma^*p$ c.m.\ frame and by
their angle $\theta$ with respect to the proton flight direction, their
pseudorapidity $\eta=-\ln[\tan(\theta/2)]$, and their energy $E=x_E E_p$ in
the laboratory frame.
They were detected within the acceptance cuts $p_T^*>2.5$~GeV or 3.5~GeV,
$5^\circ<\theta<25^\circ$, and $x_E>0.01$.
The DIS phase space was restricted to the kinematic regime defined by
$0.1<y<0.6$ and $2<Q^2<70$~GeV$^2$.
The cross section was measured differentially in $p_T^*$
\cite{Adloff:1999zx,Aktas:2004rb}, $\eta$ \cite{Adloff:1999zx}, $x_E$
\cite{Aktas:2004rb}, and $x_B$ \cite{Adloff:1999zx,Aktas:2004rb} for various
$Q^2$ intervals, differentially in $x_E$ for various $x_B$ intervals
\cite{Aktas:2004rb}, and differentially in $Q^2$ \cite{Adloff:1999zx}.
The differential cross sections $d\sigma^{\pi^0}/dp_T^*$,
$d\sigma^{\pi^0}/d\eta$, $d\sigma^{\pi^0}/dx_E$, $d\sigma^{\pi^0}/dx_B$, and
$d\sigma^{\pi^0}/dQ^2$ presented in Refs.~\cite{Adloff:1999zx} (open circles)
and \cite{Aktas:2004rb} (solid circles) are compared with our LO (dashed
histograms) and NLO (solid histograms) predictions in
Figs.~\ref{fig:p}--\ref{fig:q}, respectively.
In Figs.~\ref{fig:p}, \ref{fig:e}, \ref{fig:x}(a), and \ref{fig:b}(a), the
upper three frames refer to the $Q^2$ intervals $2<Q^2<4.5$~GeV$^2$,
$4.5<Q^2<15$~GeV$^2$, and $15<Q^2<70$~GeV$^2$.
In Fig.~\ref{fig:x}(b), the upper three frames refer to the $x_B$ intervals
$0.000042<x_B<0.0002$, $0.0002<x_B<0.001$, and $0.001<x_B<0.0063$.
In Fig.~\ref{fig:b}(b), the upper three frames refer to the $Q^2$ intervals
$2<Q^2<8$~GeV$^2$, $8<Q^2<20$~GeV$^2$, and $20<Q^2<70$~GeV$^2$.
In all figures, the minimum-$p_T^*$ cut is $p_T^*>2.5$~GeV, expect for
Fig.~\ref{fig:b}(b), where it is $p_T^*>3.5$~GeV.
In Figs.~\ref{fig:p}--\ref{fig:q}, the shaded bands indicate the theoretical
uncertainties of the NLO predictions due to the $\xi$ variation described
above.
The $K$ factors, defined as the NLO to LO ratios of our default predictions,
are shown in the downmost frames of Figs.~\ref{fig:p}--\ref{fig:q}.

\begin{figure}[t]
\begin{center}
\includegraphics[width=14cm]{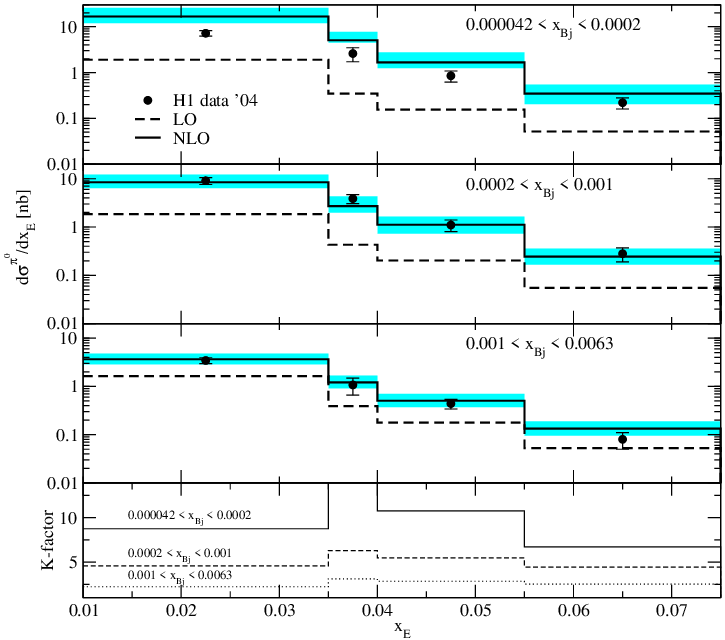}\\
{Figure~\ref{fig:x}: (b) Same as in Fig.~\ref{fig:x}(a), but for
$2<Q^2<70$~GeV$^2$ and $0.00042<x_B<0.0002$ (first frame), $0.0002<x_B<0.001$
(second frame), or $0.001<x_B<0.0063$ (third frame).}
\end{center}
\end{figure}

\begin{figure}[t]
\begin{center}
\includegraphics[width=14cm,angle=270]{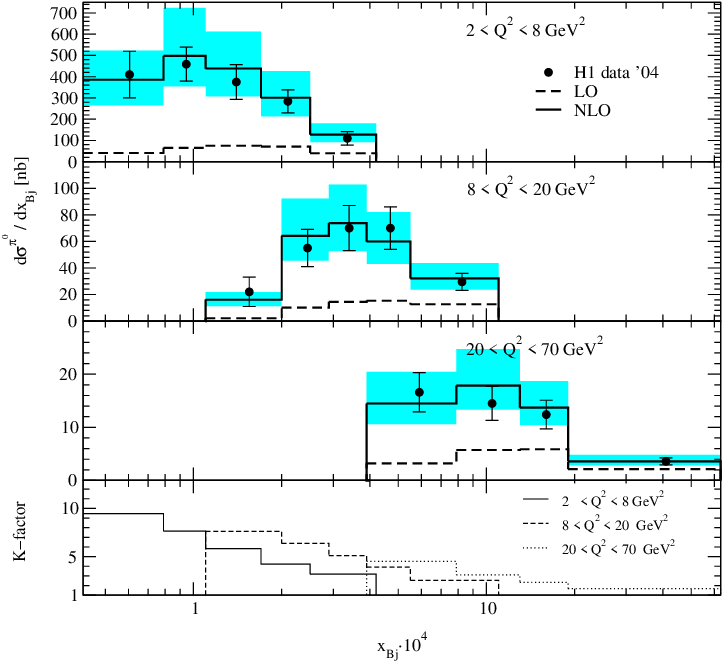}\\
{Figure~\ref{fig:b}: (b) Same as in Fig.~\ref{fig:b}(a), but for
$p_T^*>3.5$~GeV and $2<Q^2<8$~GeV$^2$ (first frame), $8<Q^2<20$~GeV$^2$
(second frame), or $20<Q^2<70$~GeV$^2$ (third frame).}
\end{center}
\end{figure}

We observe from Figs.~\ref{fig:p}--\ref{fig:q}, that the H1 data generally
agree with our NLO predictions within errors, while they significantly
overshoot our default LO predictions.
Indeed, the $K$ factors always exceed unity and even reach one order of
magnitude at low values of $p_T^*$, $Q^2$, or $x_B$.
Not only do the LO predictions disagree with the H1 data in their
normalizations, but they also exhibit deviating shapes.
On the other hand, under the effect of asymptotic freedom, the $K$ factors
approach unity for increasing values of $\mu_r$, i.e.\ for increasing values
of $p_T^*$ and/or $Q^2$.

There is an obvious explanation for the sizeable $K$ factors at low values of
$\mu_r$ in terms of the different kinematic constraints at LO and NLO.
The LO processes~(\ref{born1})--(\ref{born3}) are $2\to2$, and their cross
sections are sensitive to collinear singularities only as $p_T^*\to0$.
By contrast, processes~(\ref{real1})--(\ref{real8}) contributing to the real
NLO correction are $2\to3$, so that collinear configurations can also arise for
finite values of $p_T^*$.
After mass factorization of the corresponding collinear singularities, the
finite remainders can be sizeable, leading to large NLO corrections.
A similar line of reasoning was presented in Ref.~\cite{Aurenche:2003by}.

Unfortunately, the theoretical uncertainties in our NLO predictions due to
$\xi$ variation are rather sizeable, especially at low values of $p_T^*$,
$Q^2$, or $x_B$, where the $K$ factors themselves are abnormally large.
This is partly related to the opening of new partonic production channels at
NLO, which are still absent at LO, namely those of Eqs.~(\ref{real4}),
(\ref{real6}), and (\ref{real8}).
Obviously, a reduction in $\xi$ dependence can only be expected to happen at
next-to-next-to-leading order (NNLO), which is beyond the scope of this work.

Besides the freedom in the choice of the renormalization and factorization
scales, there are other sources of theoretical uncertainty, including the
variations of the PDF and FF sets.
However, in view of the considerable spread in cross section induced by the
moderate $\xi$ variations described above, we conclude that the residual
sources of theoretical uncertainty are of minor importance.
Furthermore, we must bear in mind that the factorization theorem itself is
only valid up to terms of ${\mathcal O}(\Lambda_{\text QCD}^2/(p_T^*)^2)$,
which may become large in the low-$p_T^*$ range.

\begin{figure}[t]
\begin{center}
\includegraphics[width=12cm,angle=270]{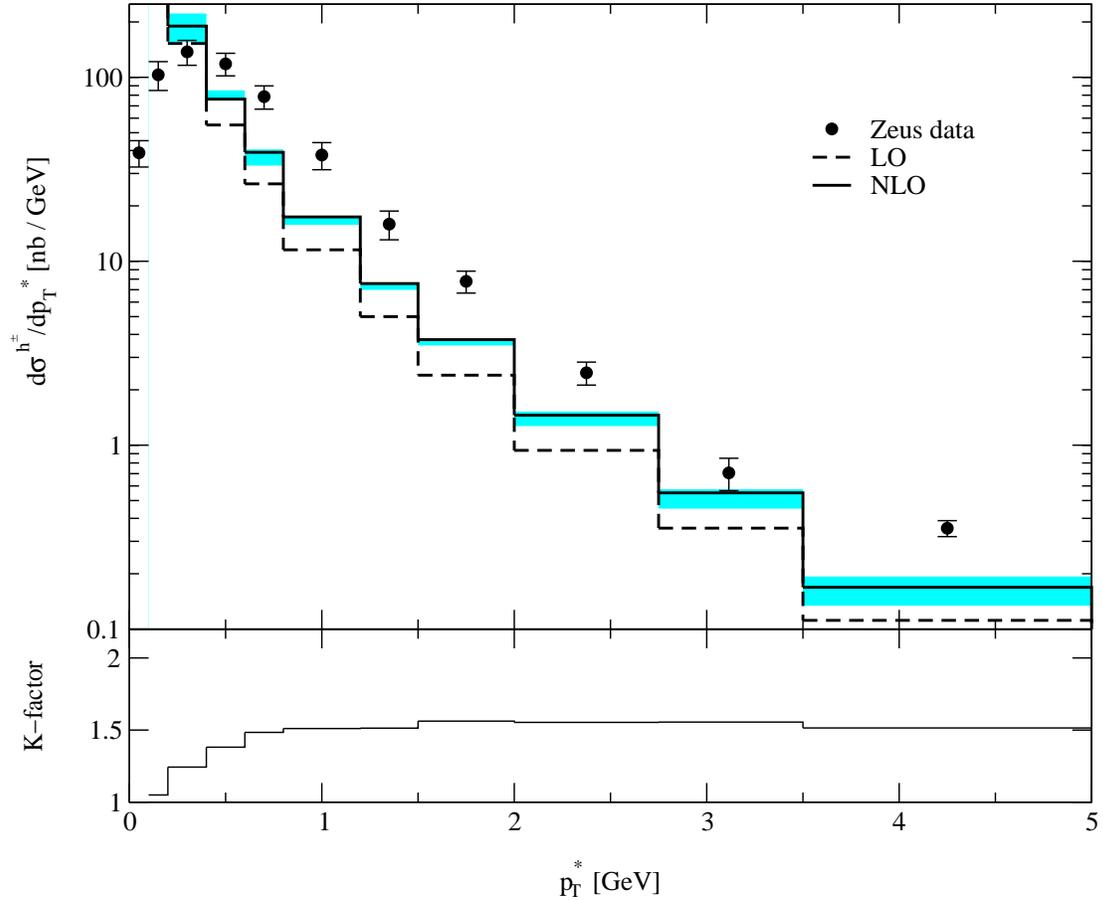}
\caption{Differential cross section $d\sigma^{h^\pm}/dp_T^*$ (in pb/GeV) of
$e^-p\to e^-h^\pm+X$ in DIS with $10<Q^2<160$~GeV$^2$ and $75<W<175$~GeV at
HERA with $E_e=26.7$~GeV and $E_p=820$~GeV for charged hadrons with
$x_F>0.05$.
ZEUS data \cite{Derrick:1995xg} (solid circles) are compared with our default
LO (dashed histograms) and NLO (solid histograms) predictions including
theoretical uncertainties due to $\xi$ variation (shaded bands).
The $K$ factor (lower frame) is also shown.}
\label{fig:zp}
\end{center}
\end{figure}

We now turn to the ZEUS data on charged hadrons \cite{Derrick:1995xg}, which
were produced in DIS of electrons with energy $E_e=26.7$~GeV on protons with
energy $E_p=820$~GeV in the laboratory frame, giving $\sqrt S=296$~GeV, during
the 1993 running period and correspond to an integrated luminosity of
0.55~pb$^{-1}$.
They refer to the DIS phase space defined by $10<Q^2<160$~GeV$^2$ and
$75<W<175$~GeV, where $W$ is the $\gamma^*p$ invariant mass, with
$W^2=(P+q)^2=(1-x_B)yS$, and come as multiplicities differential in $p_T^*$
or Feynman's $x$ variable $x_F=2p_L^*/W$, where $p_L^*=p_T^*\sinh\eta^*$ is
the projection of the hadron three-momentum onto the flight direction of the
virtual photon in the $\gamma^*p$ c.m.\ frame, and normalized to the total
number of DIS events.
Unfortunately, the $x_F$ distribution of the multiplicity includes charged
hadrons with $p_T^*$ values down to zero, while our NLO analysis is only valid
for finite values of $p_T^*$, so that a comparison is impossible.
However, a comparison is feasible for the $p_T^*$ distribution
$(1/N_\text{evt})dN_\text{had}/dp_T^*$ \cite{Derrick:1995xg}, which includes
charged hadrons with $x_F>0.05$.
The differential cross section $d\sigma^{h^\pm}/dp_T^*$ may be obtained using
the conversion formula \cite{wolf}
\begin{equation}
\frac{1}{\sigma_\text{tot}^\text{DIS}}\,\frac{d\sigma^{h^\pm}}{dp_T^*}
=\frac{1}{N_\text{evt}}\,\frac{dN_\text{had}}{dp_T^*},
\end{equation}
where $\sigma_\text{tot}^\text{DIS}$ is the total cross section in the DIS
regime specified above,
\begin{equation}
\sigma_\text{tot}^\text{DIS}=\int_{Q_\text{min}^2}^{Q_\text{max}^2}dQ^2
\int_{W_\text{min}}^{W_\text{max}}dW\frac{d^2\sigma^\text{DIS}}{dQ^2\,dW}.
\end{equation}
At LO, we have \cite{pdg}
\begin{equation}
\frac{d^2\sigma^\text{DIS}}{dQ^2\,dW}
=4\pi\alpha^2\frac{W}{Q^6}x_B[1+(1-y)^2]F_2^\gamma(x_B,Q^2),
\end{equation}
where $x_B=Q^2/(Q^2+W^2)$ and $y=(Q^2+W^2)/S$.
Using the parameterization \cite{wolf}
\begin{equation}
F_2^\gamma(x_B,Q^2)=c_1\left(\frac{1}{x_B}\right)^{c_2+c_3\ln(1+Q^2/Q_0^2)},
\end{equation}
where $Q_0^2=0.4$~GeV$^2$, $c_1=0.2030\pm0.0086$, $c_2=0.0727\pm0.0046$, and
$c_3=0.0448\pm0.0012$, obtained from a fit to ZEUS data, we thus find
$\sigma_\text{tot}^\text{DIS}=(35.4\pm2.1)$~nb assuming the errors on $c_1$,
$c_2$, and $c_3$ to be statistically independent.
This nicely agrees with the result $\sigma_\text{tot}^\text{DIS}=33.9$~nb
obtained in the parton model of QCD, where \cite{pdg}
\begin{equation}
F_2^\gamma(x_B,Q^2)=x_B\sum_{i=1}^{n_f}e_{q_i}^2
\left[F_{q_i}^p(x_B,Q^2)+F_{\overline{q}_i}^p(x_B,Q^2)\right],
\end{equation}
using set CTEQ6L1 \cite{Pumplin:2002vw} of proton PDFs with $n_f=5$.
For consistency, we use the ZEUS result for $\sigma_\text{tot}^\text{DIS}$ to
convert the ZEUS data for $(1/N_\text{evt})dN_\text{had}/dp_T^*$
\cite{Derrick:1995xg}.
The result for $d\sigma^{h^\pm}/dp_T^*$ thus obtained (solid circles) is
compared with our LO (dashed histogram) and NLO (solid histogram) predictions
in Fig.~\ref{fig:zp} (upper frame).
As in Figs.~\ref{fig:p}--\ref{fig:q}, the shaded band indicates the theoretical
uncertainty in the NLO prediction due to the $\xi$ variation described above,
and the $K$ factor is also shown (lower frame).
Again, our NLO prediction leads to a better description of data than our LO 
one.
Here, the $K$ factor takes more moderate values than under H1 kinematic
conditions, being of order 1.5 or below.
As explained above, our LO and NLO predictions break down in the limit
$p_T^*\to0$.
This drawback can be fixed by the resummation of multiple parton radiation,
as demonstrated in Ref.~\cite{Nadolsky:2000ky} on the basis of the LO result.

\begin{figure}[t]
\begin{center}
\includegraphics[width=14cm]{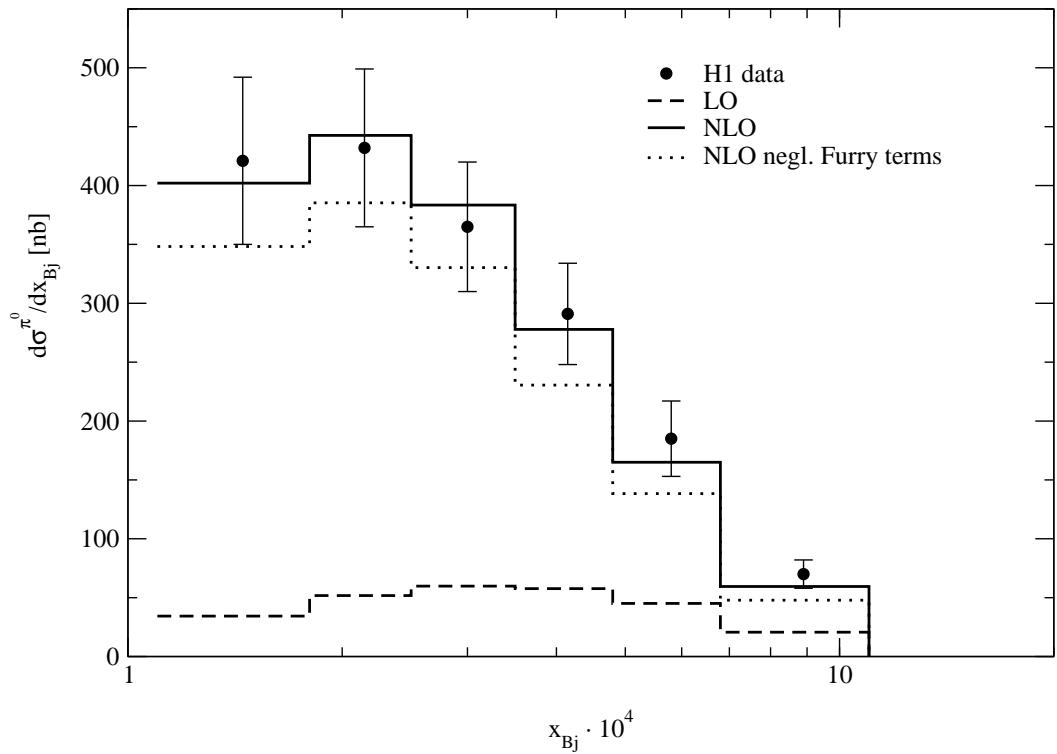}
\caption{Same as in (a) the second frame of Fig.~\ref{fig:b}(a) and (b) the
first frame of Fig.~\ref{fig:zp}, but also including our default NLO
predictions with the Furry terms turned off (dotted histograms).
For clarity, the theoretical uncertainties due to $\xi$ variation are
omitted.}
\label{fig:F}
\end{center}
\end{figure}

In Section~\ref{sec:ana}, we explained why the Furry terms do not vanish in
our case, in contrast to inclusive jet production in DIS \cite{Catani:1996gg}.
It is interesting to investigate their importance quantitatively.
To this end, we reconsider the differential cross sections
$d\sigma^{\pi^0}/dx_B$ for $0.1<y<0.6$, $4.5<Q^2<15$~GeV$^2$, $p_T^*>2.5$~GeV,
$5^\circ<\theta<25^\circ$, and $x_E>0.01$ and $d\sigma^{h^\pm}/dp_T^*$ for
$10<Q^2<160$~GeV$^2$, $75<W<175$~GeV, and $x_F>0.05$, which we already studied
in the second frame of Fig.~\ref{fig:b}(a) and the first frame of
Fig.~\ref{fig:zp}, respectively, and turn off the Furry terms in our default
NLO prediction.
The results are shown together with our default LO and NLO predictions and the
H1 \cite{Aktas:2004rb} and ZEUS \cite{Derrick:1995xg} data in
Figs.~\ref{fig:F}(a) and (b), respectively.
We observe that the Furry terms are very important.
In Fig.~\ref{fig:F}(a), they account for roughly 20\% of the NLO correction,
while, in Fig.~\ref{fig:F}(b), they practically exhaust the latter.

\begin{figure}[t]
\begin{center}
\includegraphics[width=10cm,angle=270]{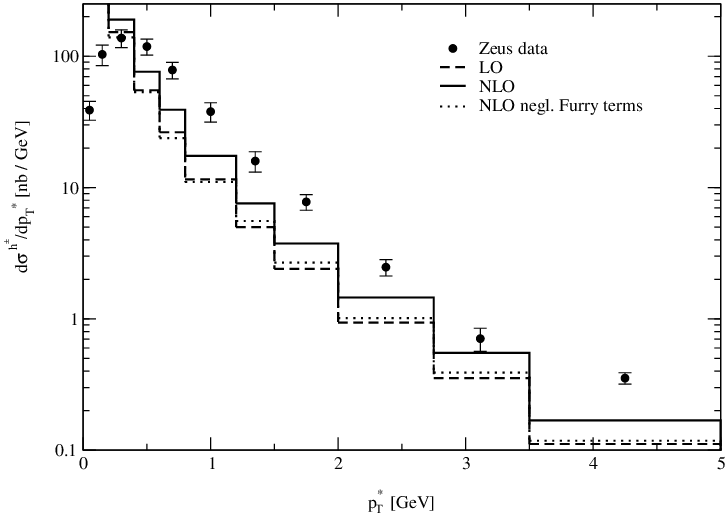}\\
{Figure~\ref{fig:F}: Continued.}
\end{center}
\end{figure}

We expect our fixed-order predictions to break down in three extreme kinematic
regimes corresponding to the limits (i) $Q^2\to0$; (ii) $\theta\to0$ or,
equivalently, $\eta\to\infty$ or $x_F\to-1$; and (iii) $x_B\to0$. 
Case (i) corresponds to the photoproduction limit, in which the resolved-photon
contribution gains importance, especially at small values of $p_T^*$ and/or
$\theta$.
Case (ii) is related to the possibility that the observed hadron $h$
originates from the proton remnant, so that the notion of fracture functions
is invoked.
Case (iii) is expected to correspond to the realm of BFKL \cite{BFKL}
dynamics, although it is unclear precisely where the onset of the latter is
supposed to be located.
Our analysis is puristic in the sense that resolved virtual photons, fracture
functions, and BFKL dynamics are disregarded, so as to test their actual
relevance in the confrontation of the QCD-improved parton model with the
experimental situation of
Refs.~\cite{Adloff:1999zx,Aktas:2004rb,Derrick:1995xg}.
Let us now scrutinize these issues.
Doing this, however, we have to bear in mind that the theoretical uncertainty
in our NLO predictions due to the arbitrariness in the choice of the
unphysical scales is particularly large in these corners of phase space, so
that any conclusions are likely to be premature prior to the advent of a full
NNLO analysis. 
From Fig.~\ref{fig:q}, we observe that our NLO prediction for
$d\sigma^{\pi^0}/dQ^2$ tends to undershoot the H1 data \cite{Adloff:1999zx} in
the low-$Q^2$ range, so that there is indeed some room for a resolved-photon
contribution.
Similar conclusions were reached in Ref.~\cite{Fontannaz:2004qv}.
On the other hand, we see from Fig.~\ref{fig:e} that the H1 data for
$d\sigma^{\pi^0}/d\eta$ \cite{Adloff:1999zx} significantly exceed our NLO
prediction in the very forward region, i.e.\ in the rightmost $\eta$ bin, for
low values of $Q^2$.
In fact, for $2<Q^2<4.5$~GeV$^2$, the measured $\eta$ distribution exhibits a
plateau in the upper $\eta$ range, whereas the NLO prediction is rapidly
suppressed by the shrinkage of the available phase space for increasing value
of $\eta$.
This plateau might be partly caused by $\pi^0$ mesons originating from the
remnant jet, which contaminate the proper data sample.
Such events cannot be described within our puristic NLO QCD framework.
Finally, thanks to the support from the Furry terms, we find in
Figs.~\ref{fig:b}(a) and (b) satisfactory overall agreement between our NLO
prediction for $d\sigma^{\pi^0}/dx_B$ and the H1 data \cite{Aktas:2004rb} down
to the lowest $x_B$ values considered.
A similar conclusion can be drawn from Fig.~\ref{fig:x}(b) for
$d\sigma^{\pi^0}/dx_E$ in the low-$x_B$ bin
$4.2\times10^{-5}<x_B<2\times10^{-4}$.
This suggest that, in the case of light-hadron inclusive production in DIS at
HERA, the influence of the BFKL dynamics is likely to be still feeble for
$x_B\agt4.2\times10^{-5}$.

\section{Conclusion}
\label{sec:con}

We analytically calculated the cross section for the inclusive
electroproduction of single hadrons with finite transverse momenta via
virtual-photon exchange at NLO in the QCD-improved parton model, with $n_f$
massless quark flavours, on the basis of the collinear-factorization theorem.
We worked in the $\overline{\text{MS}}$ renormalization and factorization
scheme and handled the IR singularities using the dipole subtraction
formalism \cite{Catani:1996vz}.
As for the virtual correction, we reproduced the result of 
Ref.~\cite{Graudenz:1993tg}.
As for the real correction, we established agreement with
Ref.~\cite{Catani:1996gg}, up to the Furry terms, which vanish upon
phase-space integration in the case of single-jet inclusive electroproduction
considered in Ref.~\cite{Catani:1996gg}, but yield a finite contribution in
the case under consideration here.

Using nonperturbative FFs recently extracted from data of $e^+e^-$
annihilation \cite{Kniehl:2000fe}, we provided theoretical predictions for the
production of $\pi^0$ mesons in the forward region and of charged hadrons in
the current-jet region, and compared them in all possible ways with H1
\cite{Adloff:1999zx,Aktas:2004rb} and ZEUS \cite{Derrick:1995xg} data,
respectively.
Specifically, we considered cross section distributions in $p_T^*$, $\eta$,
$x_E$, $x_B$, and $Q^2$.

We found that our LO predictions always significantly fell short of the HERA
data and often exhibited deviating shapes.
However, the situation dramatically improved as we proceeded to NLO, where our
default predictions, endowed with theoretical uncertainties estimated by
moderate unphysical-scale variations, led to a satisfactory description of the
HERA data in the preponderant part of the accessed phase space.
In other words, we encountered $K$ factors much in excess of unity, except
towards the regime of asymptotic freedom characterized by large values of
$p_T^*$ and/or $Q^2$.
This was unavoidably accompanied by considerable theoretical uncertainties.
Both features suggest that a reliable interpretation of the HERA data
\cite{Adloff:1999zx,Aktas:2004rb,Derrick:1995xg} within the QCD-improved
parton model ultimately necessitates a full NNLO analysis, which is presently
out of reach, however.
For the time being, we conclude that the successful comparison of the HERA
data with our NLO predictions provides a useful test of the universality and
the scaling violations of the FFs, which are guaranteed by the factorization
theorem and are ruled by the DGLAP evolution equations, respectively.

Significant deviations between the HERA data and our NLO predictions only
occurred in certain corners of phase space, namely in the photoproduction
limit $Q^2\to0$, where resolved virtual photons are expected to contribute,
and in the limit $\eta\to\infty$, where fracture functions are supposed to enter
the stage.
Both refinements were not included in our analysis.
Interestingly, distinctive deviations could not be observed towards the lowest
$x_B$ values probed, which indicates that the realm of BFKL \cite{BFKL}
dynamics has not actually been accessed yet.

\section*{Note added}

After finalizing this manuscript, a paper has appeared which also reports on a
NLO analysis of the inclusive electroproduction of single hadrons with finite
transverse momenta \cite{dfs} reaching conclusions similar to ours.

\section*{Acknowledgements}

We thank G\"unter Wolf for a clarifying communication \cite{wolf} regarding
the extraction of $d\sigma^{h^\pm}/dp_T^*$ from Ref.~\cite{Derrick:1995xg} and
Elisabetta Gallo for drawing Ref.~\cite{Breitweg:1999nt} to our attention.
We are grateful to Michael Klasen for his colaboration at the initial stage of
this work, to Michael Spira for a beneficial communication regarding the
application of the dipole subtraction formalism to the case of two tagged
partons, and to Dominik St\"o{}ckinger for helpful discussions.
This work was supported in part by the Bundesministerium f\"ur Bildung und
Forschung through Grant No.\ 05~HT4GUA/4 and by the Deutsche
Forschungsgemeinschaft through Grant No.\ KN~365/3-1.

\appendix

\section{LO results}
\label{AppLO}

In this appendix, we list the LO expressions for $H_T^{ab}$ and $H_L^{ab}$ in
Eq.~(\ref{eqlepton}) pertaining to processes~(\ref{born1})--(\ref{born3}),
with $ab=qq,qg,gq$, respectively.
We have
\begin{eqnarray}
H_T^{qq}&=&16\pi\alpha_sC_Fe_q^2\frac{1+(1-x-z)^2}{(1-x)(1-z)},
\nonumber\\
H_L^{qq}&=&8\pi\alpha_sC_Fe_q^2Q^2\frac{z}{x},
\nonumber\\
H_T^{qg}&=&16\pi\alpha_sC_Fe_q^2\frac{1+(x-z)^2}{(1-x)z},
\nonumber\\
H_L^{qg}&=&8\pi\alpha_sC_Fe_q^2Q^2\frac{1-z}{x},
\nonumber\\
H_T^{gq}&=&\frac{16\pi\alpha_sN_cC_Fe_q^2}{N_c^2-1}\,
\frac{1-2x(1-x)-2z(1-z)}{z(1-z)},
\nonumber\\
H_L^{gq}&=&\frac{16\pi\alpha_sN_cC_Fe_q^2Q^2}{N_c^2-1}\,\frac{1-x}{x},
\end{eqnarray}
where $N_c=3$ is the number of quark colours and $C_F=(N_c^2-1)/(2N_c)=4/3$ is
the eigenvalue of the Casimir operator in the fundamental representation of
the QCD gauge group SU($N_c$).

\section{Real correction: Furry terms}
\label{AppFurry}

In this appendix, we list the NLO expressions for $H_T^{ab}$ and $H_L^{ab}$ in
Eq.~(\ref{eqlepton}) that originate from hindered Furry cancellations in the
squared matrix elements of processes~(\ref{real5}) and (\ref{real6}), with
$ab=qq$, and of process~(\ref{real8}), with $ab=qq^\prime$.
We denote the four-momenta of the second and third final-state quarks by $p_b$
and $p_c$, respectively, and introduce the invariants
$s_{ij}=p_i\!\cdot\!p_j$, where $i,j=a,b,c,d$ with $i\ne j$.
For given $q$, $p_a$ and $p_b$, we need to integrate over $p_c$, while $p_d$ is
fixed through four-momentum conservation to be $p_d=q+p_a-p_b-p_c$.
We work in the coordinate frame defined in Fig.~\ref{coordinates} and
parameterize $p_c$ as
\begin{equation}
p_c^\mu=\frac{x_c}{2}\sqrt{\frac{Q^2}{x(1-x)}}
(1,\cos\alpha\sin\beta,\sin\alpha\sin\beta,\cos\beta),
\end{equation}
where $\alpha$ and $\beta$ are the azimuthal and polar angles, respectively.
Then, we have
\begin{equation}
H_{T,L}^{F,ab}=\frac{2}{\pi}\alpha_s^2C_Fe_ae_bQ^2\frac{1-x}{x}
\int_{1-z}^zdx_c\int_0^{2\pi}d\alpha\int_{-1}^1d\cos\beta\,h_{T,L}^{F,ab},
\end{equation}
where
\begin{eqnarray}
h_T^{F, qq} &=&
        \frac{1}{s s_{ab} s_{cd}}\,\frac{1}{(p_c - q)^2}
    \bigg\{ s_{ac}^2 s_{bd} - s_{ab}^2 s_{cd} 
      - s_{ac} [ s_{ad} s_{bc} + s_{bd} ( s_{bd} + s_{cd} ) ]   
\nonumber\\
&&{}+ s_{ab} [ s_{ad} s_{bc} - 2 s_{bc} s_{bd} + 
         s_{ac} ( s_{bd} - s_{cd} )  - 
         s_{bd} s_{cd} + s_{cd}^2 ]  
\nonumber\\
&&{}- s_{ad} ( 2 s_{bd} s_{cd} + 
         s_{bc} ( 3 s_{bd} + s_{cd} ) ]  
         \bigg\}
\nonumber\\
        &&{}-(p_a \leftrightarrow -p_b)
\nonumber\\
        &&{}+(p_b \leftrightarrow p_c,\quad p_a \leftrightarrow -p_d)
\nonumber\\
        &&{}-(p_c \leftrightarrow p_d)
\nonumber\\
        &&{}+(p_b \leftrightarrow p_c)
\nonumber\\
        &&{}-(p_a \rightarrow -p_c,\quad p_b \rightarrow -p_a,\quad
 p_c \rightarrow p_b)
\nonumber\\
        &&{}+(p_a \leftrightarrow -p_d)
\nonumber\\
        &&{}-(p_b \rightarrow p_c,\quad p_c \rightarrow p_d,\quad
 p_d \rightarrow p_b),
\\
h_L^{F, qq} &=&
        \frac{1}{s_{ab} s_{ac} s_{bd} s_{cd}}
        \bigg\{ 
        \frac{1}{(p_b-q)^2(p_d-q)^2}
       s_{ac} s_{bd}
       ( s_{ac}^2 - s_{ab} s_{ad} - s_{ac} s_{ad} )
\nonumber\\
&&{}\times
        ( - s_{ad} s_{bc}  + 
         s_{ac} s_{bd} - s_{ab} s_{cd} ) 
\nonumber\\
&&{}+\frac{1}{(p_c-q)^2}
\bigg[\frac{1}{(p_d-q)^2}
s_{ab} s_{cd}
( s_{ab}^2 - s_{ab} s_{ad} - s_{ac} s_{ad} )
( s_{ab} s_{cd} - s_{ad} s_{bc} - s_{ac} s_{bd} )  
\nonumber\\     
&&{}-\frac{1}{(p_b-q)^2}
( s_{ac} s_{bd} - s_{ad} s_{bc} + s_{ab} s_{cd} )
( s_{ac} s_{ad} s_{bd} ( s_{ac} - s_{ad} ) 
+ s_{ab}^2 s_{cd} ( s_{ac} + s_{ad} )
\nonumber\\
&&{}+ s_{ab} ( s_{ac}^2 s_{bd} - s_{ad}^2 s_{cd} ) ) \bigg] \bigg\},
\\
h_T^{F,qq^\prime} &=&\frac{1}{s s_{ab} s_{cd}}\,\frac{1}{(p_c - q)^2}
\bigg\{ s_{ac}^2 s_{bd} - s_{ab}^2 s_{cd} 
      - s_{ac} [ s_{ad} s_{bc} + 
         s_{bd} ( s_{bd} + s_{cd} ) ]   
\nonumber\\
&&{}+ s_{ab} [ s_{ad} s_{bc} - 2 s_{bc} s_{bd} + 
         s_{ac} ( s_{bd} - s_{cd} ) - 
         s_{bd} s_{cd} + s_{cd}^2 ]  
\nonumber\\
&&{}- s_{ad} [ 2 s_{bd} s_{cd} + 
         s_{bc} ( 3 s_{bd} + s_{cd} ) ]  
         \bigg\}
\nonumber\\
        &&{}-(p_a \leftrightarrow -p_b)
\nonumber\\
        &&{}+(p_b \leftrightarrow p_c,\quad p_a \leftrightarrow -p_d)
\nonumber\\
        &&{}-(p_c \leftrightarrow p_d)
\\
h_L^{F, qq^\prime} &=&
        \frac{1}{s_{ab} s_{cd}}\,\frac{1}{(p_b-q)^2}
        \bigg\{\frac{1}{(p_d-q)^2}
( s_{ac}^2 - s_{ab} s_{ad} - s_{ac} s_{ad} )
( - s_{ad} s_{bc} + s_{ac} s_{bd} - s_{ab} s_{cd} ) 
\nonumber\\
&&{}-\frac{1}{(p_c-q)^2}
[ s_{ab} s_{ac} + s_{ad} ( s_{ac} - s_{ad} ) ]
( s_{ac} s_{bd} -  s_{ad} s_{bc} + s_{ab} s_{cd} )  
\bigg\}.
\end{eqnarray}

\end{document}